\documentclass[acmtog]{acmart} 

\AtBeginDocument{%
  \providecommand\BibTeX{{%
    \normalfont B\kern-0.5em{\scshape i\kern-0.25em b}\kern-0.8em\TeX}}}
\author{Maryam Riahi}
\affiliation{
  \institution{North Carolina State University}
  \streetaddress{Dept. Computer Science, Campus Box 8206}
  \city{Raleigh} \state{NC} \postcode{27695-8206} 
  \country{USA}
}
\email{bwatson@ncsu.edu}\author{Benjamin Allen Watson}
\affiliation{
  \institution{North Carolina State University}
  \streetaddress{Dept. Computer Science, Campus Box 8206}
  \city{Raleigh} \state{NC} \postcode{27695-8206} 
  \country{USA}
}
\email{bwatson@ncsu.edu}
\setcopyright{acmcopyright}
\copyrightyear{2018}
\acmYear{2018}
\acmDOI{10.1145/1122445.1122456}
\usepackage{float}
\usepackage{multirow}
\usepackage{xcolor}
\setcopyright{acmcopyright}
\acmJournal{PACMCGIT}
\acmYear{2021} \acmVolume{4} \acmNumber{1} \acmArticle{} \acmMonth{5} \acmPrice{15.00}\acmDOI{10.1145/3451269}

\begin{document}

\title{Am I Playing Better Now? The Effects of G-SYNC in 60Hz Gameplay}


\renewcommand{\shortauthors}{Riahi and Watson, et al.}

\begin{abstract}
G-SYNC technology matches formerly regular display refreshes to irregular frame updates, improving frame rates and interactive latency. In a previous study of gaming at the 30Hz frame rates common on consoles, players of Battlefield 4 were unable to discern when G-SYNC was in use, but scored higher with G-SYNC and were affected emotionally. We build on that study with the first examination of G-SYNC’s effects at the 60Hz frame rate more common in PC gaming and on emerging consoles. Though G-SYNC's effects are less at 60Hz than they were at 30Hz, G-SYNC can still improve the performance of veteran players, particularly when games are challenging. G-SYNC's effects on emotion and experience were limited.
\end{abstract}




\maketitle

\section{Introduction}

Latency affects video gameplay \cite{claypool2010latency,watson2019effects}, which can be quite quickly paced. Because fixed display refresh rates began to limit latency improvements, graphics hardware (GPU) and display manufacturers introduced adaptive synchronization (ASync) technology, which allows GPUs to avoid waits when delivering imagery to displays. NVIDIA claims that the tear-free immersion of its G-SYNC technology will beat gamers' expectations \cite{GSYNCUlt35:online}. Similarly, AMD promises to ``puts an end to choppy gameplay'' with FreeSync \cite{RadeonF22:online}.

Watson et al. \cite{watson2019effects} recently showed that G-SYNC can improve gaming performance and experience, in the 30Hz frame rate gameplay typical of many gaming consoles. However, PC players often have higher frame rates on their systems \cite{pcrefresh} and prefer them if possible. Gameplay performance at higher frame rates is also important in esports \cite{charleer2018real, hamari2017esports}, which is already a billion-dollar industry \cite{esport2019} with hundreds of millions of viewing hours \cite{esportview2019}, and continues to grow. We report here on two new studies of the effects of G-SYNC on player performance and experience at 60Hz frame rates. 

\subsection{Contributions}

With these two experiments:

\begin{itemize}
    \item We describe the first study of G-SYNC's effects on gameplay at 60Hz frame rates, common on PCs and new gaming consoles.
    
    \item We learn that G-SYNC's effects on performance and experience are more limited at 60Hz than at 30Hz, particularly for novice gamers and easier gameplay.
    
    \item Nevertheless, G-SYNC at 60Hz still enables veteran gamers to improve their performance during challenging gameplay.
    
    \item We introduce a new application of the implicit association test (IAT) \cite{egloff2002predictive,greenwald2000using} to measure gaming engagement. 
\end{itemize}

\section{Adaptive Synchronization}

For decades, display refresh rates were higher than rendered frame rates. But as GPUs improved and frame rates overtook refresh rates, GPUs more often had to wait for the display to refresh to avoid image tearing, increasing latency. ASync eliminates such waits by synchronizing refreshes to frame updates, removing not only delays caused when GPUs are faster than displays, but also those caused when GPUs are slower. In this case, the GPU misses a display refresh, and the display repeats the last frame. This introduces ``jitter'', unusually long frames that annoy gamers. Note that as average frame and refresh rates increase (e.g. from 30 to 60Hz), the delays introduced by waits and misses decrease, so that ASync's elimination of them brings fewer improvments. NVIDIA calls its ASync technology G-SYNC, AMD's more open alternative is FreeSync.

Studying the effects of ASync on user experience is a new line of research. Poth et al. \cite{poth2018ultrahigh} used G-SYNC to maximize temporal resolution by setting GPU frame rate just above the display refresh rate. Though temporal improvement was only several milliseconds, viewers recognized letters better, demonstrating G-SYNC's value outside of gaming. Within gaming, Watson et al. \cite{watson2019effects} studied the effects of G-SYNC on players performance and experience in a 30Hz gaming environment. Players were unable to identify gaming sessions using G-SYNC, even though their game scores improved with it, and their emotions were affected by it. As measured by both scores and surveys of emotion, G-SYNC interacted with playing time, game content, and the general gaming expertise of players.

\section{Latency's Effects on Gaming}
The effects of frame rate and latency on human performance and experience have been researched for decades, the bulk of it concerned with applications other than gaming \cite{claypool2006latency,chen2007review}. Frame rate (temporal sampling) is the number of images displayed per second. Latency (delay or lag) is the time between the user’s input (e.g. mouse click) and the system’s resulting output \cite{deber2015much}, but the exact definition can vary across researchers. Frame rate is a significant component of latency, but other system characteristics (such as the input device) also influence it. Here we concentrate first on latency's impacts on performance, then on experience. In both cases, gaming-related latency research is limited, so we briefly broaden the discussion.

\subsection{Effects on Performance}
The impact of latency on gameplay performance depends on the nature of the game's tasks. Claypool et al. \cite{claypool2019game} studied the effect of delay on selecting a moving target with a mouse in a custom game called ``Juke''. Delay's impact depended on the target's angle and frequency of motion. Dick et al. \cite{dick2005analysis} analyzed the effects of network latency, jitter, and player skill on perception in four different multiplayer games, and learned that effects depended heavily on the game being played. In their two studies, Long et al. \cite{Long:2018:CME:3242671.3242678} varied latency while measuring player performance and experience, and found that latency's effects depended in particular on game speed, with faster-paced games affected by smaller latencies. In esports, Spjut et al. \cite{spjut2019latency} found that latency affected performance more on complex tasks, while the frame rate's effects were minimal.

Although the reported thresholds at which latency has effects vary widely (23 to 500ms), in real-time games like first-person shooters (FPSs) performance is harmed by even small amounts of lag. FPS games are fast-paced, with accurate aiming and quick reaction crucial for good performance. FPS players are aware of latency, with studies showing that 100ms latency is noticeable, and 200ms is annoying \cite{beigbeder2004effects}. Others found that methods compensating for latency work well with FPS games \cite{ivkovic2015quantifying,lee2018enhancing}. For example, Quake IV uses built-in algorithms to compensate for packet loss \cite{wattimena2006predicting}. 

Outside of gaming, even slight delays in touch interfaces (e.g. trackpads, tablets and phones) can affect human performance. Janzen et al. \cite{janzen201460} found that 30Hz frame rates   (33.3 ms frame times) made it difficult for users to select moving targets using laptop touchpads. In a study by Jota et al. \cite{jota2013fast}, latencies as low as 25ms affected dragging on direct-touch surfaces like tablets. Deber et al. \cite{deber2015much} followed up this work, finding that direct-touch devices (e.g., phones and tablets) are more sensitive to latency than indirect-touch devices (e.g., mice and touchpads). Long et al. \cite{long2019effects} studied game input devices (mouse, touchscreen, gamepad, and drawing tablet), and learned that latency affects each device differently. Only the touchscreen was unaffected by latency in targeting stationary objects. In motion-tracked systems such as virtual reality, Ellis et al. \cite{ellis2013exploring} used machine learning to recognize poses more rapidly, reducing interactive latency.

\subsection{Effects on Experience}
A few projects have found that low latency can improve game experience. Long et al. varied latency during gameplay, and found changes in enjoyment, competence, frustration, and performance \cite{Long:2018:CME:3242671.3242678}. Claypool et al. \cite{claypool2016effects} varied game speed and delay while players used a mouse to select a moving target. Delay harmed experience more than performance. Hohlfeld et al. \cite{hohlfeld2016insensitivity} found that delay did not affect the experience of novice Minecraft players but speculated that delay might affect expert players. 

Latency aside, research suggests that gaming experience has complex relationships to the characteristics of gamers, and the games they are playing \cite{liu2009dynamic,alharthi2018playing}. In particular, the relationship between game difficulty and experience has been studied extensively. For example, Frommel et al. \cite{frommel2018emotion} used dynamic difficulty adjustment (DDA) to respond to players' self-reports of boredom and frustration, camouflaged in dialog with non-player characters (NPCs) and found that it was more effective than simpler DDA control schemes. Others suggest predicting emotion using machine learning, processing elements of player behavior \cite{frommel2018towards,roohi2019recognizing}. It can be tempting to conclude that experiencing positive emotion implies a good gaming experience, but research \cite{cole2015emotional,bopp2018odd,bopp2016negative,gowler2019horror} suggests that this relationship is more complex, with negative emotions often creating better experience.

 \begin{table}
 \centering
  \label{participants}
  \caption{Participants vs. expertise and ordering \\ in the first experiment.}
  \label{tab:Table1}
  \begin{tabular}{cccrrrcrrr}
    \toprule
    
    & & & \multicolumn{3}{c}{Gaming Expertise} & \\
    
    \noalign{\smallskip}
    
    & & &  < 5 & > 5 & \textit{total}  \\
    
    \midrule
    
    \multirow{2}{*}{G-SYNC} & first  & & 9 & 8 & \textit{17} \\ 
                            & second & & 7 & 8  & \textit{15} \\
    & \textit{total} & & \textit{16} & \textit{16} & \textit{32} \\

    \bottomrule
  \end{tabular}

\end{table}

\section{Experiment 1: G-SYNC Within Participants} \label{sec:within}

 Watson et al. \cite{watson2019effects} showed G-SYNC can improve gaming performance and experience in 30Hz frame rate systems, and found intriguing relationships between G-SYNC's effects and the traits of both games and gamers. We wished to study the same questions in a more PC-centric and future-proof 60Hz gaming context. To do so, we carefully designed a 60Hz PC gaming environment and introduced an implicit measure of experience, the Implicit Association Test (IAT) \cite{greenwald1998measuring}.

\subsection{Participants}

32 students participated in our experiment, all undergraduates or graduates in Computer Science. We recruited them using email announcements and compensated them with extra credit in their course or with \$10 Amazon gift cards. Table \ref{tab:Table1} shows how we assigned participants to different between-participant groups.

\begin{figure*}
  \includegraphics[width=2.15\columnwidth]{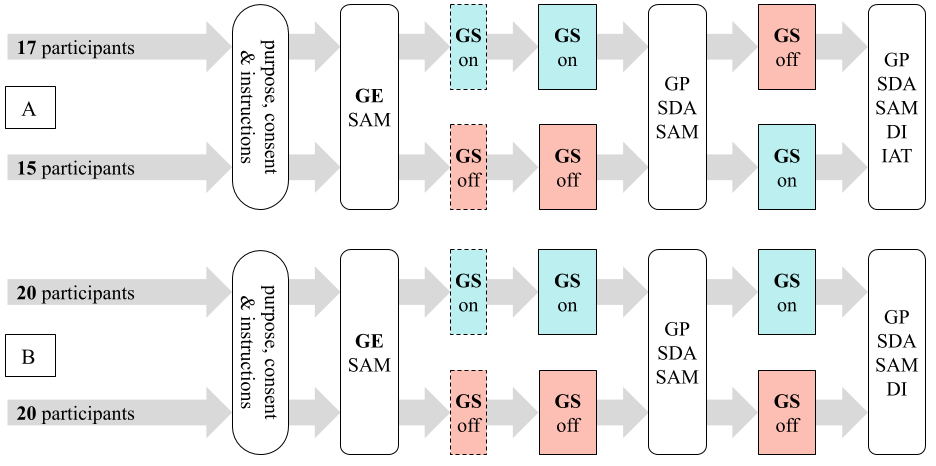}
  \caption{\textit{A} shows procedure for our first, G-SYNC within participants experiment (Section \ref{sec:within}); and \textit{B} for our between participants experiment (Section \ref{sec:between}). Bold abbreviations are independent variables, others are dependent measures (see Sections \ref{subsub:independents} and \ref{subsub:dependents}). Participants proceed from left to right: through intro (oval), measurement (white), gameplay (colored) and practice (dashed) stages. Blue shows \textit{G-SYNC } on, red \textit{G-SYNC} off. }
  ~\label{fig:figure1}
\end{figure*}

\subsection{Design}
We used a 3-factor mixed design ($2 GS \times 2 SN \times 2 GE $). Figure \ref{fig:figure1} shows our independent variables and dependent measures, and how they related to the experimental procedure.

\subsubsection{Independent variables}  \label{subsub:independents}

To study the effect of G-SYNC, all participants played a session on Battlefield 4 (B4) with and without \textit{G-SYNC} (\textit{GS}). Thus \textit{GS} was within subjects. 

As participants begin and then continue playing a game, their familiarity with the game grows, while the game typically becomes more challenging. We investigated the relationship of these effects to G-SYNC by dividing gameplay into two \textit{sessions} (\textit{SN}). Our \textit{session} variable was also within-subjects, although it was not completely crossed with \textit{G-SYNC}: participants did not play both with and without \textit{G-SYNC} in each \textit{session}. We considered alternative designs that did not couple learning with game difficulty but rejected them because the repetitive play they required would alter the introductory gaming experience we were measuring.

All participants had played FPS games. We named those who had played FPS games for at least 5 years \textit{veterans}, others were \textit{novices}. Thus our \textit{gaming expertise} variable (\textit{GE}) was between participants. In Watson et al.'s work \cite{watson2019effects}, veterans often performed better and had different experiences than novice players. We considered measures of expertise with more recency, such as hours played per week (more or less than 6). However, this measure was correlated with years of FPS gaming among our participants (novices: 69\% < 6 hours; veterans: 75\% > 6). Only one participant listed B4 as one of the games they actively play, while the others listed different FPS games.

\subsubsection{Dependent measures} \label{subsub:dependents}

We discuss our dependent measures in the order participants encountered them, as shown in Figure \ref{fig:figure1}.

\textit{Self-Assessment Manikin (SAM)}. SAM is a simple, pictorial survey for assessing a person\textquotesingle s emotional state \cite{bradley1994measuring}, based on the three-dimensional model described by Mehrabian and Russell \cite{mehrabian1974approach}. In that model, the pleasure dimension (\textit{SAM-P}) places emotion on an axis labeled with adjective pairs such as ``unhappy-happy,'' ``annoyed-pleased,'' and ``unsatisfied-satisfied.'' Arousal (\textit{SAM-A}) is labeled with pairs such as  ``relaxed-stimulated,'' ``calm-excited,'' and ``sluggish-frenzied.'' Finally, dominance (\textit{SAM-D}) corresponds to ``controlled-controlling,'' ``influenced-influential,'' and ``cared for-in control.'' Participants described their emotions using SAM three times: before, between, and after gameplay sessions. We used the difference between succeeding \textit{SAM} assessments to assess the emotional change after gameplay sessions, which resulted in a between - before SAM difference and an after - between difference.

\textit{Game Performance (GP)}. To measure player performance, we used B4's single-player scoring system. After each gaming session, we recorded the new points earned (points at end of session less points at beginning). B4 awards points for each kill: regular 100 points, headshot 125, multi- 100 points each, adrenaline 150, melee 125, streak 150, and squad 50. We sought other measures less sensitive to variation in game difficulty and easier to generalize to other games but did not find any. We examine this in more detail in section \ref{subsection:limitations}.

\textit{Subjective Duration Assessment (SDA)}. SDA is an implicit measure of subjective difficulty introduced by Czerwinski et al. \cite{czerwinski2001subjective}, who argued that users underestimate actual task time when the task feels easy, and overestimate task time when it feels difficult. Bederson also argues that SDA is a measure of flow \cite{bederson}, or full involvement in an activity, which is closely related to difficulty \cite{nakamura2014concept}. Participants estimated the number of minutes they had played after each gaming session. Each session's \textit{SDA} measure was then the participant estimate less the actual number of minutes they had played. 

\textit{Display Identification (DI)}. To measure the perceptibility of G-SYNC, after their last gaming session, we asked participants, ``Which monitor do you think was equipped with G-SYNC, the improved display?'' They then identified the gaming session with the display they found better. The \textit{DI} measure is the ratio of participants who correctly identified the \textit{G-SYNC} session.

\textit{Implicit Association Test (IAT)}. As an implicit measure of gameplay engagement, we used the IAT. Implicit measures do not rely primarily on conscious thought and are useful when researchers are concerned about cognitive biases. For example, social desirability bias drives participants to give the answers they think researchers want. Since its introduction in 1998 \cite{greenwald1998measuring}, the IAT has been widely adopted, improved, and validated \cite{greenwald2003understanding, lane2007understanding, greenwald2009understanding}. The IAT measures the subconscious association between an ``attribute'' construct and a characterizing ``target'' construct. For example, the IAT is often used to measure association between race (the target) and goodness (the attribute). To determine associative strength, IAT participants classify examples (phrases or imagery) of both constructs, while they view two pairs of those constructs (Figure \ref{fig:figure2}). If participants implicitly agree with the pair, they will classify more quickly. We used the IAT to measure the association between engagement (attribute) and gaming session (target)  after all gameplay was complete. 

\subsection{Apparatus}

B4 is supported by NVIDIA drivers, is popular among FPS gamers \cite{battlefield4}, is visually rich and requires rapid reaction. As the game used by Watson et al. \cite{watson2019effects}, it also allows us to extend their work.

In configuring our experimental system, we sought to realize a B4 frame rate of 60Hz. Higher frame rates exist in gaming, but this is twice the frame rate used by Watson et al. \cite{watson2019effects}, is commonly cited by gamers as a desirable goal \cite{sarkar2014frame,gamingscan}, matches a common refresh rate setting in displays, and approaches the peak rate at which display flicker can be perceived \cite{landis1954determinants}. We did not measure input-to-display latency, but varied only latency's frame rate component.

To examine configurations, we varied PC and GPU hardware, display refresh rates, and B4 settings including maximum frame rate and level of detail. Our secondary goals were a meaningful improvement frame time mean and standard deviation by G-SYNC, and avoiding configurations unusual for gamers. To measure frame rates, we used the Fraps tool \cite{FRAPSgam40:online}, without video recording. B4 is now 6 years old, so our final system is modest and used an Intel Core i7-870 2.93GHz CPU with 8GB RAM operated by 64-bit Windows, an NVIDIA GeForce GTX 1050 Ti GPU, an Acer XB240H-A G-SYNC monitor displaying 1920$\times$1080 pixels and using a 144Hz refresh rate 
and double buffering, with B4 set to a maximum frame rate of 62Hz and ``high'' level of detail (B4's second-highest setting). This gave a mean frame rate of 57.683 Hz (\(\sigma\) = 2.22) without G-SYNC and a mean frame rate of 59.467 Hz (\(\sigma\) = 1.83) with G-SYNC. When G-SYNC was off, VSync was on. Players used mouse and keyboard with B4. 

\begin{figure}
  \centering
  \includegraphics[width=0.98\linewidth]{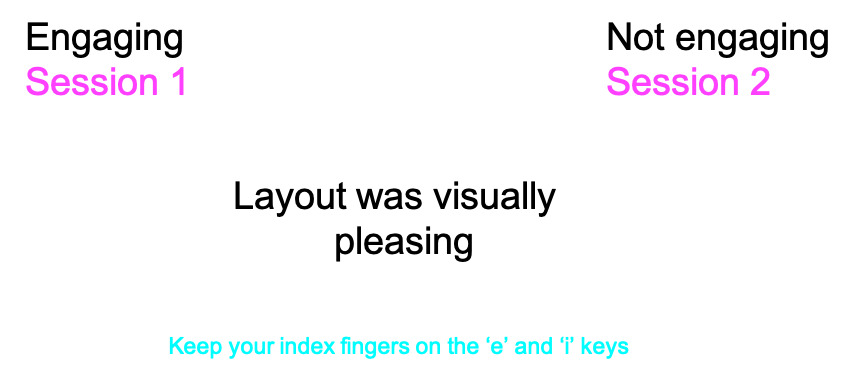}
  \caption{An example IAT screen. Participants press ``E'' if they classify the center example with the left pair, and ``I'' if they classify with the right.}
  \Description{An example IAT screen}
  ~\label{fig:figure2}
\end{figure}

We used Meade's FreeIAT software \cite{meade2009freeiat}, which uses the updated IAT procedure in \cite{greenwald2003understanding}. As examples of the gaming session construct, we used phrases identifying a part of the B4 mission, such as ``protecting Irish,'' ``helicopter,'' ``safe house,'' etc. As examples of the engagement construct, we used phrases such as ``I was absorbed,'' ``It was demanding,'' ``It was confusing'' drawn from Wiebe et al.'s gaming engagement survey \cite{wiebe2014measuring}. Participants input their classification of examples by pressing E for the left-hand construct pair, or I for the right-hand pair (Figure \ref{fig:figure2}). As an index of associative strength, the IAT reports the difference between average classification times in two blocks with reversed pairings. In our experiment, negative differences showed more engagement with the second session, and positive differences more engagement with the first. 

\subsection{Procedure and Task}

We randomly determined whether participants would begin gameplay with or without G-SYNC. They then moved through seven experimental stages (Figure \ref{fig:figure1}), which required 44.5 minutes on average. We discuss these stages below.

\subsubsection{Before gameplay}

When they arrived, we thanked participants for their help and told them that they could halt the experiment at any time. We informed them that we were studying NVIDIA's G-SYNC in gaming, adding that they would use G-SYNC in one of their gaming sessions, but they would not know which until the experiment's end. They would be describing their emotions and thoughts about their gameplay and should be honest in their responses since there is no ``wrong'' answer and their responses would be anonymous. Participants gave informed consent for their participation, with the assurance that their data would be protected.

We then surveyed participants to find which \textit{gaming expertise} group they belonged to and to assess their emotional state using the SAM. This pre-gameplay SAM allowed us to control for individual differences in participant emotion that did not result from gameplay. 

Since only one of the participants was an active B4 player (but all had played FPS games), they gained familiarity with B4 in an initial practice game session using B4's ``BAKU'' mission in the ``easy'' mode. Participants halted the mission when they felt comfortable with the game or had been playing for 10 minutes. When they died, B4 would restart them from the last checkpoint. With experimenter guidance, participants practiced employing keys and mouse to move and use weapons, and began learning B4's maps, weapons and help system. Participants beginning gameplay with G-SYNC on practiced with G-SYNC on, those beginning without practiced with it off.

\subsubsection{First gaming session and measures}

Participants then played in the experiment's first gaming session, continuing B4's ``BAKU'' mission in the ``easy'' mode. Because we measured both performance and experience, we did not give players a particular gameplay goal. The mission asked players to find teammates, acquire weapons, exit a school, and fight through enemies. The session ended when players reached an open field. Participants played with G-SYNC if they were in that random group, without otherwise. They continued until reaching the field (\(\mu = 9.5\) minutes), or for 15 minutes. We then recorded participants' B4 score for the GP measure, surveyed participants to obtain the SAM measure, and asked participants to estimate playing time for the SDA measure (participants were unaware of any time limit). We then toggled G-SYNC.

\subsubsection{Second gaming session and measures}

This session's portion of the mission was more difficult, with more enemy soldiers and vehicles. Participants began in the same open field that ended the first session and then attempted to cross an open field under intense enemy fire to reach a large drainage pipe. Play halted when they reached the pipe, or 15 minutes had passed. All participants used the full 15 minutes. We again recorded B4 scores for our GP measure, surveyed participants for our SAM measure, and asked for a playing time estimate for the SDA measure. We also asked participants to identify the gaming session that used G-SYNC for our DI measure, and they performed the IAT procedure to measure engagement. We then told participants which session used G-SYNC. 

\begin{figure}
\centering
  \includegraphics[width=1\columnwidth]{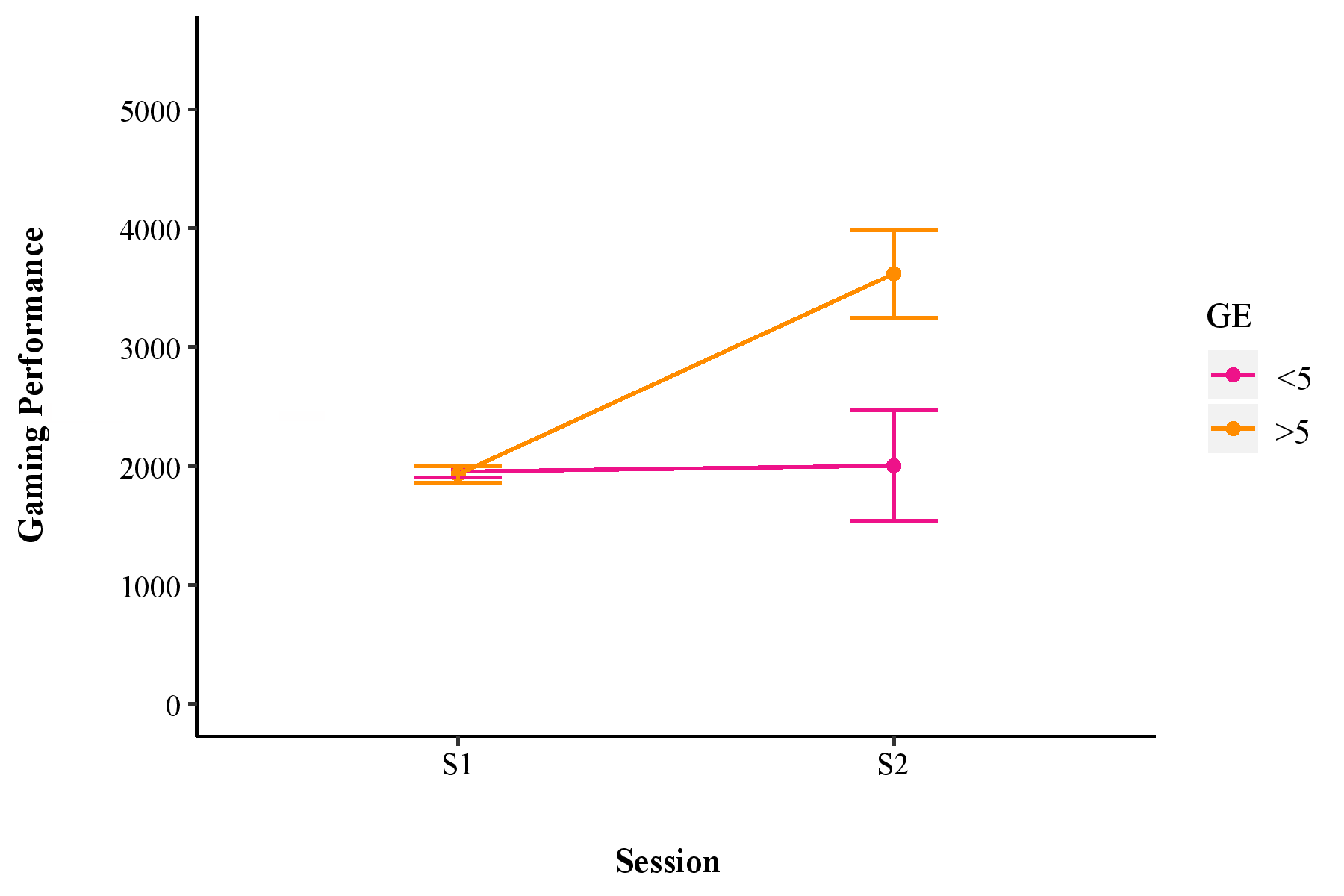}
  \caption{Experiment 1: Mean scores from the within participants study, grouped by \textit{session} and \textit{gaming expertise}. Bars show standard error.}~\label{fig:figure3}
\end{figure}

\subsection{Hypotheses}
Our hypotheses were that:

\subsubsection{\textit{G-SYNC} would be \textit{imperceptible}}
The average difference between the G-SYNC on and G-SYNC off conditions was only about 3Hz (centered around 60Hz), which translates to a \textasciitilde0.8ms difference in frame time. Such temporal frequency differences are at perceptual limits \cite{watson201164}, and these latency differences are well below reported limits (e.g. \cite{Annett:2014:LWG:2619648.2619677}). Moreover, with larger temporal differences, Watson et al. \cite{watson2019effects} found that G-SYNC's effect was imperceptible.

\subsubsection{\textit{G-SYNC} would \textit{improve performance slightly}}
Research shows that lower latencies improve task performance \cite{jota2013fast, deber2015much}. In Watson et al. \cite{watson2019effects}, G-SYNC also improved gaming performance; but in our study, G-SYNC's temporal effects were more moderate. 

\subsubsection{\textit{G-SYNC} would \textit{improve experience slightly}.}
Research is limited but suggests that experience can respond strongly to frame rate and latency, and interacts with gaming, content, and expertise. However, Watson et al. \cite{watson2019effects} found modest effects on experience, and this study's latency improvements were more moderate.

\subsection{Results}
We report significant (p < .05) and marginal (p < .1) effects on perceptibility, performance, and experience. For most of our results, we performed a 3-factor between-participants ANOVA with \textit{G-SYNC} (\textit{GS}), \textit{session} (\textit{SN}), and \textit{gaming expertise} (\textit{GE}) as factors. This between-participants analysis simplifies comparisons with our second experiment. Moreover, because \textit{GS} and \textit{SN} were not fully crossed, we could not use them as within factors in a 3-factor mixed ANOVA. We compared interaction means with contrasts.

\subsubsection{Perceptibility}
Participants identified the G-SYNC display correctly 53\% of the time. This was not significantly different from chance (50\%) (\(\chi^2 \)(1,32) = 0.03, p = .85). Participants could not reliably discern sessions with G-SYNC from sessions without it.

\subsubsection{Performance}
\textit{G-SYNC} did not significantly effect B4 scores. Table \ref{tab:Table2} shows significant effects. \textit{Gaming expertise} had a significant effect, with veteran's scores 40\% higher than novices' scores (<5: \(\mu = 1978.1\), standard error (SE) = 231.1; >5: \(\mu = 2774.3\), SE = 238.8). \textit{Session} also had a significant effect, with scores in the second session 45\% higher than those in the first (S1: \(\mu = 1942.3\), SE = 41.7; S2: \(\mu = 2810.1\), SE = 326.8). Moreover, \textit{GE} and \textit{SN} interacted significantly (Figure \ref{fig:figure3}): while novices' scores did not improve in the second session (S1: \(\mu = 1953.1\), SE = 48.0; S2: \(\mu = 2003.1\), SE = 467.3; p = .9975, d = 0.038), veterans' scores did (S1: \(\mu = 1931.4\), SE = 69.7; S2: \(\mu = 3617.1\), SE = 369.0; p = .0015, d = 1.587).

\subsubsection{Experience}
\textit{G-SYNC}'s only effect was on game engagement. IAT methodology reports association strength as classification time differences using Cohen's d effect size \cite{greenwald2003understanding}, with 0.8 or larger a ``strong'' association, 0.5 or larger ``moderate'' and 0.2 or larger ``slight''. Participants found the second \textit{session} more engaging, but this association did not reach slight strength (d = -0.149). However, for those using \textit{GS} in the second session, this association reached a slight, significant strength (d = -0.251, t(13)=-2.86, p=0.014), showing that \textit{GS} helped participants feel slightly more engaged in the second session. Among those using \textit{GS} in the first session, any association of engagement with session was absent (d = -0.070). 

We found no significant effects on SDA, nor on SAM valence and dominance. But \textit{session} did affect SAM arousal significantly (Table \ref{tab:Table2}), with participants gaining more excitement in the first session than the second (S1: \(\mu = -0.688\), SE = 0.158; S2: \(\mu = -0.156\), SE = 0.169). An interaction of \textit{SN} with \textit{GE} showed that \textit{SN}'s effect was focused on novices: while they became more excited during the first session, they became slightly calmer during the second (S1: \(\mu = -0.813\), SE = 0.228; S2: \(\mu = 0.250\), SE = 0.194; p = 0.01, d = -1.257). On the other hand, veterans gained excitement in both sessions (S1: \(\mu = -0.563\), SE = 0.223; S2: \(\mu = -0.563\), SE = 0.241; p = 1, d = 0).

\subsection{Discussion}
We first examine our hypotheses about \textit{G-SYNC}'s effects, and continue with an examination of \textit{session} and \textit{expertise}'s impacts.

\subsubsection{Hypotheses about \textit{G-SYNC's} effects}
We verified our \textit{perceptibility} hypothesis: participants could not reliably identify the gaming session using G-SYNC. This is not surprising, considering that G-SYNC was also imperceptible in Watson et al. \cite{watson2019effects} and that G-SYNC had less temporal impact at our 60Hz mean than at Watson et al.'s 30Hz. We could not confirm our \textit{performance} hypothesis, since G-SYNC did not affect performance. However, we were correct in expecting that G-SYNC's effect would be less than it was in Watson et al., again likely due to reduced temporal impact. Our \textit{experience} hypothesis met with mixed results. As expected, G-SYNC's effects were fewer than in Watson et al., but G-SYNC did not affect the SAM emotion measures at all. We were correct in predicting that G-SYNC would have no or minor effects on the SDA measure of subjective difficulty. Finally, our expectation that G-SYNC would affect IAT engagement slightly proved correct, at least for those using G-SYNC in the second session.

\subsubsection{\textit{Session} and \textit{expertise} effects}
Several effects reassure us that our measures are sensitive to participant performance and experience, despite \textit{G-SYNC}'s limited impact. Veterans outscored novices in the second session when they knew the game better and had more scoring chances. This may explain why only veterans continued to become more excited through the second session. Similar effects were found by Watson et al. \cite{watson2019effects}, though \textit{session} and \textit{expertise} impacted SAM valence and dominance rather than arousal.

\begin{table}
 \centering
  \caption{ANOVAs for experiment 1, within participants.}
  \label{tab:Table2}
  \begin{tabular}{llrrl}
    \toprule
    
    \textit{measure} & \textit{effects} & \textit{distribution} & \textit{p value} & \textit{$\eta^2$}   \\
    
    \midrule
    
      GP & SN & F(1,56) = 8.22 & = 0.005 & = 0.12\\
      GP & GE & F(1,56) =  6.48 & = 0.013 & = 0.103\\
      GP & SN \(\times\) GE & F(1,56) = 6.73 & = 0.012 & = 0.107 \\
      SAM-A & SN & F(1,56) =  5.35 & = 0.024 & = 0.087\\
      SAM-A & SN \(\times\) GE & F(1,56) =  5.35 & = 0.024 & = 0.087\\

    \bottomrule
  \end{tabular}

\end{table}
\begin{figure}
\centering
  \includegraphics[width=1\columnwidth]{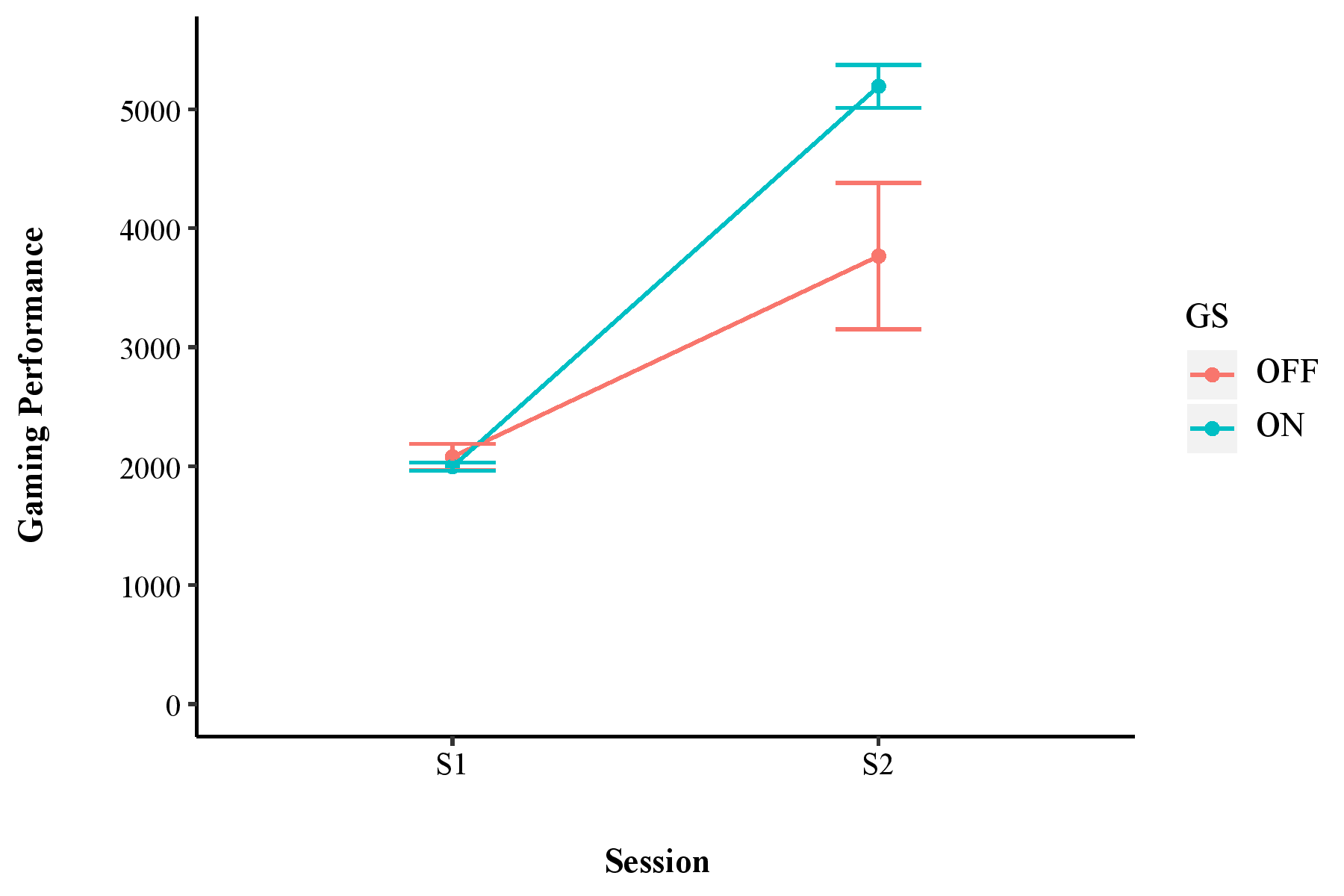}
  \caption{Experiment 2: Veteran performance in the between participants study, and its relation to \textit{G-SYNC} and \textit{session}. Effects were  significantly different from those on novices. Bars show standard error.}~\label{fig:figure4}
\end{figure}

\section{Experiment 2: G-SYNC Between Participants} \label{sec:between}

Several results from experiment 1 indicate that B4's beginning brings less challenge and experiential impact, matching a common design pattern that eases players into games, allowing them to enjoy and succeed in early gameplay, while also gaining the skill and familiarity with game \textit{and platform} that permits similar success later. 

Our first experiment's within-participant design increased statistical power and permitted participants to compare displays with and without G-SYNC. However, its mid-gameplay display switch interfered with learning of the gaming platform, and perhaps also of the game itself, just as that learning was most important. Moreover, such display technology changes are not typical of actual gameplay. For these reasons, we decided to replicate our experiment using a between-participant design, with each participant using just one display technology, either with or without G-SYNC.

The following subsections note only the differences between experiments 1 and 2.

\subsection{Participants}
To compensate for the reduced statistical power of a between-participants design, we recruited more computer science students: 20 novices (\textit{GE<5}) and 20 veterans (\textit{GE>5}).

\subsection{Design and Variables}
We again used a 3-factor mixed design (2GS×2SN×2GE). However, \textit{G-SYNC} was between-participants, with participants playing both sessions entirely with or without G-SYNC. \textit{Session} remained within-participants, and \textit{expertise} between. Figure \ref{fig:figure1} shows how this affected our procedure, with \textit{G-SYNC} on in light blue only on the top, and \textit{G-SYNC} off in pink on the bottom. Because each participant saw only one \textit{G-SYNC} setting, we could not use the IAT to measure \textit{G-SYNC}'s effect on engagement. In our \textit{perceptibility} measure, we could not ask which session used G-SYNC. Instead, we asked participants if they believed they had used an improved G-SYNC display.
\begin{table}
 \centering
  \caption{ANOVAs for experiment 2, between participants.}~\label{tab:Table3}
  \begin{tabular}{lllrl}
    \toprule
    
    \textit{measure} & \textit{effects} & \textit{distribution} & \textit{p value} & \textit{$\eta^2$}   \\
    
    \midrule
    GP & SN & F(1,36) = 43.2 & <0.001 & = 0.3\\
    GP & GE & F(1,36) =  29.2 &  <0.001 & = 0.34\\
    GP &SN \(\times\) GE & F(1,36) = 33.3 &  <0.001& = 0.25 \\
    GP & GS \(\times\) SN \(\times\) GE & F(1,36) = 6.43 &  = 0.015 & = 0.06 \\
    SAM-A & SN & F(1,36) =  5.77 & = 0.021 & = 0.11\\
    SAM-D & SN \(\times\) GE & F(1,36) =  6.24 & = 0.017 & = 0.09\\
    SDA & SN & F(1,36) =  10.8 & = 0.002 & = 0.05\\
    SDA & GS \(\times\) SN \(\times\) GE & F(1,36) =  3.5 & = 0.069 & = 0.02\\

    \bottomrule
  \end{tabular}

\end{table} 

\subsection{Procedure and Task}
We randomly assigned half of the participants into the G-SYNC on group, and the other half into the G-SYNC off group. They went through the same seven experimental stages (Figure \ref{fig:figure1}), but GSYNC was never toggled during each participant's experiment, which took 45 minutes to complete on average (SE = 1.1).

\subsection{Hypotheses}
Our hypotheses were that: 

\subsubsection{G-SYNC would be imperceptible}
Participants could not reliably perceive G-SYNC in experiment 1. In addition, in this experiment participants could not compare the two displays.

\subsubsection{G-SYNC would improve performance slightly}
 Participants in this experiment would have more exposure to G-SYNC than in experiment 1, which should increase G-SYNC's effects.  
 
\subsubsection{G-SYNC would not affect experience}
G-SYNC affected only experiment 1's IAT measure, which we could not use here.

\subsection{Results}
We performed 3-factor between-participants ANOVAs with \textit{G-SYNC}, \textit{session} and \textit{expertise} as factors. 

\subsubsection{Perceptibility}
Participants correctly identified \textit{G-SYNC} display 52\% of the time, which was not significantly different than chance (50\%) (\(\chi^2 \)(1,40) = 0.025, p = .87).

\subsubsection{Performance}
Table \ref{tab:Table3} shows significant effects on performance. \textit{G-SYNC} did not significantly affect B4 scores (on: \(\mu = 2749.4\), SE = 255.6; off: \(\mu = 2351.3\), SE = 231.0; p = 0.137, \textit{$\eta^2$} = 0.04). However, \textit{G-SYNC} interacted significantly with \textit{SN} and \textit{GE}. Novices did not improve their scores significantly in the second session, whether or not they used G-SYNC. However, veterans did improve their second session scores (Figure \ref{fig:figure4}) , with G-SYNC allowing an additional marginally significant improvement (on: \(\mu = 5192.5\), SE = 180.7; off: \(\mu = 3765.0\), SE = 614.9; p = 0.0573, d = 0.9961). While not large statistically, this improvement was meaningful in gameplay.

Scores were significantly higher in the second \textit{session}, with averages of 1900 in the first (SE = 82.7), and 3200.6 in the second (SE = 303.7). \textit{Expertise} was also significant: novices averaged 1843.1 (SE = 160.8), while veterans averaged 3257.5 (SE = 263.2). \textit{SN} and \textit{GE} had significantly related effects. While novices could not improve their scores in the second session (S1: \(\mu = 1763.8\), SE = 151.2; S2: \(\mu = 1922.5\), SE = 287.5; p = .9411, d = 0.155), veterans could (S1: \(\mu = 2036.3\), SE = 57.5; S2: \(\mu = 4478.8\), SE = 352.3; p < .0001, d = 2.164).  

\subsubsection{Experience}
\textit{G-SYNC} had no significant effects on any experiential measures. However, \textit{G-SYNC} was part of a marginally significant three-way interaction affecting SDA. Novices experienced the same level of difficulty in both sessions without G-SYNC. With it, they felt much less difficulty in the second session. Veterans' experience was nearly a mirror image: they experienced similar levels of difficulty in both sessions with G-SYNC, and without it, much less difficulty in the second session. 

\textit{Session} affected SAM arousal significantly: players gained more excitement from the first session than from the second (S1: \(\mu = -0.850\), SE = 0.146; S2: \(\mu = -0.250\), SE = 0.138). \textit{Session} and \textit{expertise} combined to affect SAM dominance significantly. Novices felt little change in control in the first session (\(\mu = 0.100\), SEM = 0.250) and a loss of control during the second (\(\mu = -0.350\), SEM = 0.196); but veterans felt a loss of control in the first session (\(\mu = -0.300\), SEM = 0.242), followed by a gain of control in the second (\(\mu = 0.450\), SEM = 0.170). \textit{Session} also affected SDA. In the first session, participants overestimated actual time by a few (\(\mu = 3.475\), SE = 0.705) minutes, experiencing difficulty. In the second, they overestimated by one minute (\(\mu = 1.150\), SE = 0.952), experiencing less difficulty.

\subsection{Discussion}
We again consider hypotheses about \textit{G-SYNC}'s effects, then examine \textit{session}'s and \textit{expertise}'s impacts.

\subsubsection{Hypotheses about \textit{G-SYNC's} effects}
We again verified our \textit{perceptibility} hypothesis: participants could not tell if they used G-SYNC. We also confirmed our slight improvement to \textit{performance} hypothesis. \textit{G-SYNC} had no broad effect on B4 scores, but it did allow veterans playing in the second, more challenging session to raise their scores by 38\%. Results largely confirmed our \textit{experience} hypothesis. \textit{G-SYNC} had no significant experiential effects, but a marginally significant three-way interaction hinted at differences in how G-SYNC affects subjective difficulty for novices and veterans.

\subsubsection{\textit{Session} and \textit{expertise} effects}
\textit{Session}'s effects on performance and experience were extensive and echoed its effects in experiment 1. Players scored more in the second session than in the first, likely due to increased scoring opportunities, and improved familiarity with the game. Players also gained more excitement and felt more difficulty in the first session than in the second.

\textit{Expertise} only affected performance, with veterans outscoring novices. However, \textit{expertise} interacted with \textit{session}, with only veterans able to increase their scoring and sense of control in the second session. The two \textit{G-SYNC} effects we found were limited to components of these interactions: veteran players in the second session were able to increase scoring when using G-SYNC, and this likely translated to their experiencing more difficulty in the second session with G-SYNC than without it.

\section{GENERAL DISCUSSION}

We begin considering our results by noting their limitations. We then examine several applied questions and less immediate implications.

\subsection{Limitations} \label{subsection:limitations}

Our research should be interpreted with caution. Most of our participants were similarly aged college students, while today's gamers are much more diverse. The only effect of G-SYNC on performance we found was focused on 20 veterans in experiment 2. A confirming experiment with more veteran gamers would be useful. 

Our \textit{expertise} variable was based on years of play. We contemplated short-term indices, but they grouped participants similarly. Our \textit{session} variable included both learning and difficulty effects, making it difficult to draw inferences about them. We considered repetitive gameplay offering such control, but opted for more typical gameplay to preserve the natural gaming experience. To measure performance, we used B4's native scoring system. This is subject to in-game change in difficulty and scoring chances, making it challenging to draw conclusions about participant learning, and to generalize our conclusions to other games. Study of high performance gaming would benefit from performance measures that allow comparison of game difficulty to learning, and different games to one another. We experimented with normalized scores and scoring percentiles; other possibilities might include Z-scores and scoring rates.

Watson et al. \cite{watson2019effects} examined 30Hz gameplay, while we examined 60Hz play. Still higher frame rates are becoming common, especially in esports.  We did not measure end-to-end (finger-to-photon) latency, limiting useful context on our experimental system. NVIDIA's new Reflex technology should make measurement simple in the future. We used Battlefield 4 in our work; a well-known example of the FPS gaming genre that is highly sensitive to latency. Nevertheless, other gaming genres such as sports and role-playing merit study. Perhaps most importantly, our study examined gameplay over a fairly short time (< 30 minutes). Typical real world play sessions can span hours, and depending on the gaming genre, players may continue their play over weeks, or even years.

\subsection{Applied Questions}
Here we offer tentative answers to complex questions.

\subsubsection{Can players tell when G-SYNC is in use?}
At higher frame rates, no. In 30Hz gameplay \cite{watson2019effects}, players could not reliably identify display with G-SYNC. The same proved true in our two experiments with 60Hz gameplay. However, extended play or high expertise may enable gamers to identify G-SYNC reliably.

\subsubsection{Does G-SYNC improve gaming performance?}
It can, as shown by Watson et al. \cite{watson2019effects}, and in a more limited way, by our work. We expect that G-SYNC's impact on gaming performance lessens as frame rates rise and latencies fall. We also found evidence suggesting that G-SYNC's impact on gaming performance depends on gamer expertise, playing time, and game content. For more detail on these relationships, see below.

\subsubsection{Does G-SYNC improve gaming experience?}
G-SYNC's effects on the initial gamer experience are limited, especially at 60Hz. In our experiments, G-SYNC increased gamer engagement slightly, and sometimes improved gamers' feelings of control and excitement. These improvements usually depended on whether gamers were novices or veterans, and what part of the game they were playing. For example in our second experiment, novices felt less control in the second part of the game, while veterans felt more control.

\subsubsection{What are the implications of G-SYNC for game designers?}
Both our work and Watson et al.'s \cite{watson2019effects} found evidence that G-SYNC can improve gaming performance. As G-SYNC and higher frame rates become more common, game designers can give players more difficult challenges. This work also found evidence suggesting that veteran players benefit more from G-SYNC than novices. Designers might be able to separate veteran from novice players by creating gaming challenges built specifically for G-SYNC-equipped systems.

\subsection{Issues and Implications} 
If we consider G-SYNC and...

\subsubsection{...Playing Time}
G-SYNC's performance benefits may increase with playing time. G-SYNC did not improve veterans' scores in our first experiment. However in our second experiment, veterans used G-SYNC for a longer time than those in the first, and it improved their scores in the second session. Increased playing time may allow gamers to learn not only about game challenges but also about G-SYNC itself, allowing gamers to exploit its potential more fully.

\subsubsection{...Expertise}
G-SYNC's performance and experiential impacts vary with gamer expertise. In our second experiment, only veterans were able to exploit G-SYNC to increase their score in the second session. In both of our experiments, measures of experience regularly varied with gamer expertise. We suspect that novices and veterans evaluate gameplay differently. For example, veterans may value gaming performance more than novices.

\subsubsection{...Game Content}
G-SYNC's performance benefits likely depend on the nature of a game's challenges. Our experiments did not explicitly control those challenges, but in our second experiment, G-SYNC's performance benefits were limited to the second session of gameplay. We expect that some of these benefits relate to the new challenges in B4's second session. In particular, tasks that require rapid responses or continuous monitoring may be particularly sensitive to the latencies that G-SYNC reduces. 

\subsubsection{...How it Helps}
Because G-SYNC's improvements in mean latency were minimal (<1ms) in our 60Hz experiments, its beneficial effect on performance was likely due to reductions in transient high latencies. Also, both novice and veteran experiences were affected by G-SYNC, but only veterans were able to realize any of G-SYNC's performance benefits. It may be that novices benefit primarily from G-SYNC's advantages for perception, which change their experience; veteran gamers also benefit from G-SYNC's advantages for action, which permit them to realize improvements in game performance. 

\subsubsection{...FreeSync}
Although we did not directly compare G-SYNC to FreeSync, we anticipate that they will have similar effects on gaming performance and experience. These ASync technologies do differ, but they have similar goals and mechanisms --- indeed, G-SYNC supports many FreeSync monitors. Nevertheless, future work should seek confirmation. 

\section{Future Work and Conclusion}
Research on ASync's impacts is sparse, leaving many promising directions for future work. Such study might address this project's limitations by introducing shorter-term indices of expertise, and increasing diversity among participants and gaming genres. It would be interesting to compare the performance and experiential impacts of ASync, VSync and no VSync (with tearing). The issues we raised in our general discussion also merit investigation, especially ASync's benefits as playing time and gamer expertise increase. In particular, study of the extreme expertise of esports athletes should be informative. More broadly, research on high-performance interfaces of all kinds is sparse. As computing technology reaches more deeply into our work and play, the need for interfaces supporting exceptional performance will continue to grow.

G-SYNC lowers latencies and increases frame rates by improving synchronization between displays and GPUs. This study builds on prior work examining G-SYNC's effects on FPS gaming performance and experience by --- for the first time --- examining its effects at the higher 60Hz frame rates common on PCs and in the latest consoles. We find that while G-SYNC's overall effects are less at 60Hz than at 30Hz, it still enables veteran gamers to improve their performance, particularly when games become more familiar and challenging.

\section{Acknowledgments}
Removed for anonymous review.

\begin{acks}
We thank all of our participants. We are also grateful to Josef Spjut and Joowhan Kim of NVIDIA Research, who gave advice throughout our project, and provided detail about G-SYNC. Many reviewers helped improve this paper. Early work on this project was performed by undergraduates supported by an NSF REU Site grant.
\end{acks}

\bibliographystyle{ACM-Reference-Format}
\bibliography{gsync60}


\begin{thebibliography}{56}


\ifx \showCODEN    \undefined \def \showCODEN     #1{\unskip}     \fi
\ifx \showDOI      \undefined \def \showDOI       #1{#1}\fi
\ifx \showISBNx    \undefined \def \showISBNx     #1{\unskip}     \fi
\ifx \showISBNxiii \undefined \def \showISBNxiii  #1{\unskip}     \fi
\ifx \showISSN     \undefined \def \showISSN      #1{\unskip}     \fi
\ifx \showLCCN     \undefined \def \showLCCN      #1{\unskip}     \fi
\ifx \shownote     \undefined \def \shownote      #1{#1}          \fi
\ifx \showarticletitle \undefined \def \showarticletitle #1{#1}   \fi
\ifx \showURL      \undefined \def \showURL       {\relax}        \fi
\providecommand\bibfield[2]{#2}
\providecommand\bibinfo[2]{#2}
\providecommand\natexlab[1]{#1}
\providecommand\showeprint[2][]{arXiv:#2}

\bibitem[\protect\citeauthoryear{Alharthi, Alsaedi, Toups, Tanenbaum, and
  Hammer}{Alharthi et~al\mbox{.}}{2018}]%
        {alharthi2018playing}
\bibfield{author}{\bibinfo{person}{Sultan~A Alharthi}, \bibinfo{person}{Olaa
  Alsaedi}, \bibinfo{person}{Zachary~O Toups}, \bibinfo{person}{Joshua
  Tanenbaum}, {and} \bibinfo{person}{Jessica Hammer}.}
  \bibinfo{year}{2018}\natexlab{}.
\newblock \showarticletitle{Playing to wait: A taxonomy of idle games}. In
  \bibinfo{booktitle}{\emph{Proceedings of the 2018 CHI Conference on Human
  Factors in Computing Systems}}. \bibinfo{publisher}{ACM},
  \bibinfo{address}{New York, NY, USA}, \bibinfo{pages}{1--15}.
\newblock


\bibitem[\protect\citeauthoryear{AMD}{AMD}{2019}]%
        {RadeonF22:online}
\bibfield{author}{\bibinfo{person}{AMD}.} \bibinfo{year}{2019}\natexlab{}.
\newblock \bibinfo{booktitle}{\emph{Radeon FreeSyncTechnology | FreeSync 2 HDR
  Games | AMD}}.
\newblock AMD.
\newblock
\urldef\tempurl%
\url{https://www.amd.com/en/technologies/free-sync}
\showURL{%
\tempurl}


\bibitem[\protect\citeauthoryear{Annett, Ng, Dietz, Bischof, and Gupta}{Annett
  et~al\mbox{.}}{2014}]%
        {Annett:2014:LWG:2619648.2619677}
\bibfield{author}{\bibinfo{person}{Michelle Annett}, \bibinfo{person}{Albert
  Ng}, \bibinfo{person}{Paul Dietz}, \bibinfo{person}{Walter~F. Bischof}, {and}
  \bibinfo{person}{Anoop Gupta}.} \bibinfo{year}{2014}\natexlab{}.
\newblock \showarticletitle{How Low Should We Go?: Understanding the Perception
  of Latency While Inking}. In \bibinfo{booktitle}{\emph{Proceedings of
  Graphics Interface 2014}} \emph{(\bibinfo{series}{GI '14})}.
  \bibinfo{publisher}{Canadian Information Processing Society},
  \bibinfo{address}{Toronto, Ont., Canada, Canada}, \bibinfo{pages}{167--174}.
\newblock
\showISBNx{978-1-4822-6003-8}
\urldef\tempurl%
\url{http://dl.acm.org/citation.cfm?id=2619648.2619677}
\showURL{%
\tempurl}


\bibitem[\protect\citeauthoryear{Bederson}{Bederson}{2004}]%
        {bederson}
\bibfield{author}{\bibinfo{person}{Benjamin~B Bederson}.}
  \bibinfo{year}{2004}\natexlab{}.
\newblock \showarticletitle{Interfaces for staying in the flow}.
\newblock \bibinfo{journal}{\emph{Ubiquity}} \bibinfo{volume}{2004},
  \bibinfo{number}{September} (\bibinfo{year}{2004}), \bibinfo{pages}{1--1}.
\newblock


\bibitem[\protect\citeauthoryear{Beigbeder, Coughlan, Lusher, Plunkett, Agu,
  and Claypool}{Beigbeder et~al\mbox{.}}{2004}]%
        {beigbeder2004effects}
\bibfield{author}{\bibinfo{person}{Tom Beigbeder}, \bibinfo{person}{Rory
  Coughlan}, \bibinfo{person}{Corey Lusher}, \bibinfo{person}{John Plunkett},
  \bibinfo{person}{Emmanuel Agu}, {and} \bibinfo{person}{Mark Claypool}.}
  \bibinfo{year}{2004}\natexlab{}.
\newblock \showarticletitle{The effects of loss and latency on user performance
  in unreal tournament 2003{\textregistered}}. In
  \bibinfo{booktitle}{\emph{Proceedings of 3rd ACM SIGCOMM workshop on Network
  and system support for games}}. ACM, \bibinfo{publisher}{ACM},
  \bibinfo{address}{New York, NY, USA}, \bibinfo{pages}{144--151}.
\newblock


\bibitem[\protect\citeauthoryear{Bopp, Mekler, and Opwis}{Bopp
  et~al\mbox{.}}{2016}]%
        {bopp2016negative}
\bibfield{author}{\bibinfo{person}{Julia~Ayumi Bopp}, \bibinfo{person}{Elisa~D
  Mekler}, {and} \bibinfo{person}{Klaus Opwis}.}
  \bibinfo{year}{2016}\natexlab{}.
\newblock \showarticletitle{Negative emotion, positive experience? Emotionally
  moving moments in digital games}. In \bibinfo{booktitle}{\emph{Proceedings of
  the 2016 CHI Conference on Human Factors in Computing Systems}}.
  \bibinfo{publisher}{ACM}, \bibinfo{address}{New York, NY, USA},
  \bibinfo{pages}{2996--3006}.
\newblock


\bibitem[\protect\citeauthoryear{Bopp, Opwis, and Mekler}{Bopp
  et~al\mbox{.}}{2018}]%
        {bopp2018odd}
\bibfield{author}{\bibinfo{person}{Julia~Ayumi Bopp}, \bibinfo{person}{Klaus
  Opwis}, {and} \bibinfo{person}{Elisa~D Mekler}.}
  \bibinfo{year}{2018}\natexlab{}.
\newblock \showarticletitle{“An Odd Kind of Pleasure” Differentiating
  Emotional Challenge in Digital Games}. In
  \bibinfo{booktitle}{\emph{Proceedings of the 2018 chi conference on human
  factors in computing systems}}. \bibinfo{publisher}{ACM},
  \bibinfo{address}{New York, NY, USA}, \bibinfo{pages}{1--12}.
\newblock


\bibitem[\protect\citeauthoryear{Bradley and Lang}{Bradley and Lang}{1994}]%
        {bradley1994measuring}
\bibfield{author}{\bibinfo{person}{Margaret~M Bradley} {and}
  \bibinfo{person}{Peter~J Lang}.} \bibinfo{year}{1994}\natexlab{}.
\newblock \showarticletitle{Measuring emotion: the self-assessment manikin and
  the semantic differential}.
\newblock \bibinfo{journal}{\emph{Journal of behavior therapy and experimental
  psychiatry}} \bibinfo{volume}{25}, \bibinfo{number}{1}
  (\bibinfo{year}{1994}), \bibinfo{pages}{49--59}.
\newblock


\bibitem[\protect\citeauthoryear{Charleer, Gerling, Guti{\'e}rrez, Cauwenbergh,
  Luycx, and Verbert}{Charleer et~al\mbox{.}}{2018}]%
        {charleer2018real}
\bibfield{author}{\bibinfo{person}{Sven Charleer}, \bibinfo{person}{Kathrin
  Gerling}, \bibinfo{person}{Francisco Guti{\'e}rrez}, \bibinfo{person}{Hans
  Cauwenbergh}, \bibinfo{person}{Bram Luycx}, {and} \bibinfo{person}{Katrien
  Verbert}.} \bibinfo{year}{2018}\natexlab{}.
\newblock \showarticletitle{Real-time dashboards to support esports
  spectating}. In \bibinfo{booktitle}{\emph{Proceedings of the 2018 Annual
  Symposium on Computer-Human Interaction in Play}}. \bibinfo{publisher}{ACM},
  \bibinfo{address}{New York, NY, USA}, \bibinfo{pages}{59--71}.
\newblock


\bibitem[\protect\citeauthoryear{Chen and Thropp}{Chen and Thropp}{2007}]%
        {chen2007review}
\bibfield{author}{\bibinfo{person}{Jessie~YC Chen} {and}
  \bibinfo{person}{Jennifer~E Thropp}.} \bibinfo{year}{2007}\natexlab{}.
\newblock \showarticletitle{Review of low frame rate effects on human
  performance}.
\newblock \bibinfo{journal}{\emph{IEEE Transactions on Systems, Man, and
  Cybernetics-Part A: Systems and Humans}} \bibinfo{volume}{37},
  \bibinfo{number}{6} (\bibinfo{year}{2007}), \bibinfo{pages}{1063--1076}.
\newblock


\bibitem[\protect\citeauthoryear{Claypool and Claypool}{Claypool and
  Claypool}{2006}]%
        {claypool2006latency}
\bibfield{author}{\bibinfo{person}{Mark Claypool} {and} \bibinfo{person}{Kajal
  Claypool}.} \bibinfo{year}{2006}\natexlab{}.
\newblock \showarticletitle{Latency and player actions in online games}.
\newblock \bibinfo{journal}{\emph{Commun. ACM}} \bibinfo{volume}{49},
  \bibinfo{number}{11} (\bibinfo{year}{2006}), \bibinfo{pages}{40--45}.
\newblock


\bibitem[\protect\citeauthoryear{Claypool and Claypool}{Claypool and
  Claypool}{2010}]%
        {claypool2010latency}
\bibfield{author}{\bibinfo{person}{Mark Claypool} {and} \bibinfo{person}{Kajal
  Claypool}.} \bibinfo{year}{2010}\natexlab{}.
\newblock \showarticletitle{Latency can kill: precision and deadline in online
  games}. In \bibinfo{booktitle}{\emph{Proceedings of the first annual ACM
  SIGMM conference on Multimedia systems}}. ACM, \bibinfo{publisher}{ACM},
  \bibinfo{address}{New York, NY, USA}, \bibinfo{pages}{215--222}.
\newblock


\bibitem[\protect\citeauthoryear{Claypool, Cockburn, and Gutwin}{Claypool
  et~al\mbox{.}}{2019}]%
        {claypool2019game}
\bibfield{author}{\bibinfo{person}{Mark Claypool}, \bibinfo{person}{Andy
  Cockburn}, {and} \bibinfo{person}{Carl Gutwin}.}
  \bibinfo{year}{2019}\natexlab{}.
\newblock \showarticletitle{Game input with delay: moving target selection
  parameters}. In \bibinfo{booktitle}{\emph{Proceedings of the 10th ACM
  Multimedia Systems Conference}}. ACM, \bibinfo{publisher}{ACM},
  \bibinfo{address}{New York, NY, USA}, \bibinfo{pages}{25--35}.
\newblock


\bibitem[\protect\citeauthoryear{Claypool, Eg, and Raaen}{Claypool
  et~al\mbox{.}}{2016}]%
        {claypool2016effects}
\bibfield{author}{\bibinfo{person}{Mark Claypool}, \bibinfo{person}{Ragnhild
  Eg}, {and} \bibinfo{person}{Kjetil Raaen}.} \bibinfo{year}{2016}\natexlab{}.
\newblock \showarticletitle{The effects of delay on game actions: Moving target
  selection with a mouse}. In \bibinfo{booktitle}{\emph{Proceedings of the 2016
  Annual Symposium on Computer-Human Interaction in Play Companion Extended
  Abstracts}}. ACM, \bibinfo{publisher}{ACM}, \bibinfo{address}{New York, NY,
  USA}, \bibinfo{pages}{117--123}.
\newblock


\bibitem[\protect\citeauthoryear{Cole, Cairns, and Gillies}{Cole
  et~al\mbox{.}}{2015}]%
        {cole2015emotional}
\bibfield{author}{\bibinfo{person}{Tom Cole}, \bibinfo{person}{Paul Cairns},
  {and} \bibinfo{person}{Marco Gillies}.} \bibinfo{year}{2015}\natexlab{}.
\newblock \showarticletitle{Emotional and functional challenge in core and
  avant-garde games}. In \bibinfo{booktitle}{\emph{Proceedings of the 2015
  annual symposium on computer-human interaction in play}}.
  \bibinfo{publisher}{ACM}, \bibinfo{address}{New York, NY, USA},
  \bibinfo{pages}{121--126}.
\newblock


\bibitem[\protect\citeauthoryear{Czerwinski, Horvitz, and Cutrell}{Czerwinski
  et~al\mbox{.}}{2001}]%
        {czerwinski2001subjective}
\bibfield{author}{\bibinfo{person}{Mary Czerwinski}, \bibinfo{person}{Eric
  Horvitz}, {and} \bibinfo{person}{Edward Cutrell}.}
  \bibinfo{year}{2001}\natexlab{}.
\newblock \showarticletitle{Subjective duration assessment: An implicit probe
  for software usability}. In \bibinfo{booktitle}{\emph{Proceedings of IHM-HCI
  2001 conference}}, Vol.~\bibinfo{volume}{2}. \bibinfo{publisher}{Springer
  Verlag}, \bibinfo{address}{London, UK}, \bibinfo{pages}{167--170}.
\newblock


\bibitem[\protect\citeauthoryear{Deber, Jota, Forlines, and Wigdor}{Deber
  et~al\mbox{.}}{2015}]%
        {deber2015much}
\bibfield{author}{\bibinfo{person}{Jonathan Deber}, \bibinfo{person}{Ricardo
  Jota}, \bibinfo{person}{Clifton Forlines}, {and} \bibinfo{person}{Daniel
  Wigdor}.} \bibinfo{year}{2015}\natexlab{}.
\newblock \showarticletitle{How much faster is fast enough?: User perception of
  latency \& latency improvements in direct and indirect touch}. In
  \bibinfo{booktitle}{\emph{Proceedings of the 33rd Annual ACM Conference on
  Human Factors in Computing Systems}}. ACM, \bibinfo{publisher}{ACM},
  \bibinfo{address}{New York, NY, USA}, \bibinfo{pages}{1827--1836}.
\newblock


\bibitem[\protect\citeauthoryear{Dick, Wellnitz, and Wolf}{Dick
  et~al\mbox{.}}{2005}]%
        {dick2005analysis}
\bibfield{author}{\bibinfo{person}{Matthias Dick}, \bibinfo{person}{Oliver
  Wellnitz}, {and} \bibinfo{person}{Lars Wolf}.}
  \bibinfo{year}{2005}\natexlab{}.
\newblock \showarticletitle{Analysis of factors affecting players' performance
  and perception in multiplayer games}. In
  \bibinfo{booktitle}{\emph{Proceedings of 4th ACM SIGCOMM workshop on Network
  and system support for games}}. ACM, \bibinfo{publisher}{ACM},
  \bibinfo{address}{New York, NY, USA}, \bibinfo{pages}{1--7}.
\newblock


\bibitem[\protect\citeauthoryear{Egloff and Schmukle}{Egloff and
  Schmukle}{2002}]%
        {egloff2002predictive}
\bibfield{author}{\bibinfo{person}{Boris Egloff} {and}
  \bibinfo{person}{Stefan~C Schmukle}.} \bibinfo{year}{2002}\natexlab{}.
\newblock \showarticletitle{Predictive validity of an implicit association test
  for assessing anxiety.}
\newblock \bibinfo{journal}{\emph{Journal of personality and social
  psychology}} \bibinfo{volume}{83}, \bibinfo{number}{6}
  (\bibinfo{year}{2002}), \bibinfo{pages}{1441}.
\newblock


\bibitem[\protect\citeauthoryear{Ellis, Masood, Tappen, LaViola, and
  Sukthankar}{Ellis et~al\mbox{.}}{2013}]%
        {ellis2013exploring}
\bibfield{author}{\bibinfo{person}{Chris Ellis}, \bibinfo{person}{Syed~Zain
  Masood}, \bibinfo{person}{Marshall~F Tappen}, \bibinfo{person}{Joseph~J
  LaViola}, {and} \bibinfo{person}{Rahul Sukthankar}.}
  \bibinfo{year}{2013}\natexlab{}.
\newblock \showarticletitle{Exploring the trade-off between accuracy and
  observational latency in action recognition}.
\newblock \bibinfo{journal}{\emph{International Journal of Computer Vision}}
  \bibinfo{volume}{101}, \bibinfo{number}{3} (\bibinfo{year}{2013}),
  \bibinfo{pages}{420--436}.
\newblock


\bibitem[\protect\citeauthoryear{Fraps}{Fraps}{2013}]%
        {FRAPSgam40:online}
\bibfield{author}{\bibinfo{person}{Fraps}.} \bibinfo{year}{2013}\natexlab{}.
\newblock \bibinfo{title}{FRAPS game capture video recorder fps viewer}.
\newblock \bibinfo{howpublished}{\url{http://www.fraps.com/}}.
\newblock
\newblock
\shownote{(Accessed on 08/29/2019).}


\bibitem[\protect\citeauthoryear{Frommel, Fischbach, Rogers, and Weber}{Frommel
  et~al\mbox{.}}{2018a}]%
        {frommel2018emotion}
\bibfield{author}{\bibinfo{person}{Julian Frommel}, \bibinfo{person}{Fabian
  Fischbach}, \bibinfo{person}{Katja Rogers}, {and} \bibinfo{person}{Michael
  Weber}.} \bibinfo{year}{2018}\natexlab{a}.
\newblock \showarticletitle{Emotion-based Dynamic Difficulty Adjustment Using
  Parameterized Difficulty and Self-Reports of Emotion}. In
  \bibinfo{booktitle}{\emph{Proceedings of the 2018 Annual Symposium on
  Computer-Human Interaction in Play}}. \bibinfo{publisher}{ACM},
  \bibinfo{address}{New York, NY, USA}, \bibinfo{pages}{163--171}.
\newblock


\bibitem[\protect\citeauthoryear{Frommel, Schrader, and Weber}{Frommel
  et~al\mbox{.}}{2018b}]%
        {frommel2018towards}
\bibfield{author}{\bibinfo{person}{Julian Frommel}, \bibinfo{person}{Claudia
  Schrader}, {and} \bibinfo{person}{Michael Weber}.}
  \bibinfo{year}{2018}\natexlab{b}.
\newblock \showarticletitle{Towards Emotion-based Adaptive Games: Emotion
  Recognition Via Input and Performance Features}. In
  \bibinfo{booktitle}{\emph{Proceedings of the 2018 Annual Symposium on
  Computer-Human Interaction in Play}}. \bibinfo{publisher}{ACM},
  \bibinfo{address}{New York, NY, USA}, \bibinfo{pages}{173--185}.
\newblock


\bibitem[\protect\citeauthoryear{game~developer EA~DICE and
  Arts}{game~developer EA~DICE and Arts}{2013}]%
        {battlefield4}
\bibfield{author}{\bibinfo{person}{Video game~developer EA~DICE} {and}
  \bibinfo{person}{Electronic Arts}.} \bibinfo{year}{2013}\natexlab{}.
\newblock \bibinfo{title}{\emph{Battlefield}}.
\newblock \bibinfo{howpublished}{Game[B4]}.
\newblock


\bibitem[\protect\citeauthoryear{Gowler and Iacovides}{Gowler and
  Iacovides}{2019}]%
        {gowler2019horror}
\bibfield{author}{\bibinfo{person}{Chad Phoenix~Rose Gowler} {and}
  \bibinfo{person}{Ioanna Iacovides}.} \bibinfo{year}{2019}\natexlab{}.
\newblock \showarticletitle{" Horror, guilt and shame"--Uncomfortable
  Experiences in Digital Games}. In \bibinfo{booktitle}{\emph{Proceedings of
  the Annual Symposium on Computer-Human Interaction in Play}}.
  \bibinfo{publisher}{ACM}, \bibinfo{address}{New York, NY, USA},
  \bibinfo{pages}{325--337}.
\newblock


\bibitem[\protect\citeauthoryear{Greenwald and Farnham}{Greenwald and
  Farnham}{2000}]%
        {greenwald2000using}
\bibfield{author}{\bibinfo{person}{Anthony~G Greenwald} {and}
  \bibinfo{person}{Shelly~D Farnham}.} \bibinfo{year}{2000}\natexlab{}.
\newblock \showarticletitle{Using the implicit association test to measure
  self-esteem and self-concept.}
\newblock \bibinfo{journal}{\emph{Journal of personality and social
  psychology}} \bibinfo{volume}{79}, \bibinfo{number}{6}
  (\bibinfo{year}{2000}), \bibinfo{pages}{1022}.
\newblock


\bibitem[\protect\citeauthoryear{Greenwald, McGhee, and Schwartz}{Greenwald
  et~al\mbox{.}}{1998}]%
        {greenwald1998measuring}
\bibfield{author}{\bibinfo{person}{Anthony~G Greenwald},
  \bibinfo{person}{Debbie~E McGhee}, {and} \bibinfo{person}{Jordan~LK
  Schwartz}.} \bibinfo{year}{1998}\natexlab{}.
\newblock \showarticletitle{Measuring individual differences in implicit
  cognition: the implicit association test.}
\newblock \bibinfo{journal}{\emph{Journal of personality and social
  psychology}} \bibinfo{volume}{74}, \bibinfo{number}{6}
  (\bibinfo{year}{1998}), \bibinfo{pages}{1464}.
\newblock


\bibitem[\protect\citeauthoryear{Greenwald, Nosek, and Banaji}{Greenwald
  et~al\mbox{.}}{2003}]%
        {greenwald2003understanding}
\bibfield{author}{\bibinfo{person}{Anthony~G Greenwald},
  \bibinfo{person}{Brian~A Nosek}, {and} \bibinfo{person}{Mahzarin~R Banaji}.}
  \bibinfo{year}{2003}\natexlab{}.
\newblock \showarticletitle{Understanding and using the implicit association
  test: I. An improved scoring algorithm.}
\newblock \bibinfo{journal}{\emph{Journal of personality and social
  psychology}} \bibinfo{volume}{85}, \bibinfo{number}{2}
  (\bibinfo{year}{2003}), \bibinfo{pages}{197}.
\newblock


\bibitem[\protect\citeauthoryear{Greenwald, Poehlman, Uhlmann, and
  Banaji}{Greenwald et~al\mbox{.}}{2009}]%
        {greenwald2009understanding}
\bibfield{author}{\bibinfo{person}{Anthony~G Greenwald},
  \bibinfo{person}{T~Andrew Poehlman}, \bibinfo{person}{Eric~Luis Uhlmann},
  {and} \bibinfo{person}{Mahzarin~R Banaji}.} \bibinfo{year}{2009}\natexlab{}.
\newblock \showarticletitle{Understanding and using the Implicit Association
  Test: III. Meta-analysis of predictive validity.}
\newblock \bibinfo{journal}{\emph{Journal of personality and social
  psychology}} \bibinfo{volume}{97}, \bibinfo{number}{1}
  (\bibinfo{year}{2009}), \bibinfo{pages}{17}.
\newblock


\bibitem[\protect\citeauthoryear{Hamari and Sj{\"o}blom}{Hamari and
  Sj{\"o}blom}{2017}]%
        {hamari2017esports}
\bibfield{author}{\bibinfo{person}{Juho Hamari} {and} \bibinfo{person}{Max
  Sj{\"o}blom}.} \bibinfo{year}{2017}\natexlab{}.
\newblock \showarticletitle{What is eSports and why do people watch it?}
\newblock \bibinfo{journal}{\emph{Internet research}} \bibinfo{volume}{27},
  \bibinfo{number}{2} (\bibinfo{year}{2017}), \bibinfo{pages}{211--232}.
\newblock
\urldef\tempurl%
\url{https://doi.org/10.1108/IntR-04-2016-0085}
\showURL{%
\tempurl}


\bibitem[\protect\citeauthoryear{Hohlfeld, Fiedler, Pujol, and Guse}{Hohlfeld
  et~al\mbox{.}}{2016}]%
        {hohlfeld2016insensitivity}
\bibfield{author}{\bibinfo{person}{Oliver Hohlfeld}, \bibinfo{person}{Hannes
  Fiedler}, \bibinfo{person}{Enric Pujol}, {and} \bibinfo{person}{Dennis
  Guse}.} \bibinfo{year}{2016}\natexlab{}.
\newblock \showarticletitle{Insensitivity to Network Delay: Minecraft Gaming
  Experience of Casual Gamers}. In \bibinfo{booktitle}{\emph{2016 28th
  International Teletraffic Congress (ITC 28)}}, Vol.~\bibinfo{volume}{3}.
  IEEE, \bibinfo{publisher}{IEEE}, \bibinfo{address}{Piscataway, NJ USA},
  \bibinfo{pages}{31--33}.
\newblock


\bibitem[\protect\citeauthoryear{Ivkovic, Stavness, Gutwin, and
  Sutcliffe}{Ivkovic et~al\mbox{.}}{2015}]%
        {ivkovic2015quantifying}
\bibfield{author}{\bibinfo{person}{Zenja Ivkovic}, \bibinfo{person}{Ian
  Stavness}, \bibinfo{person}{Carl Gutwin}, {and} \bibinfo{person}{Steven
  Sutcliffe}.} \bibinfo{year}{2015}\natexlab{}.
\newblock \showarticletitle{Quantifying and mitigating the negative effects of
  local latencies on aiming in 3d shooter games}. In
  \bibinfo{booktitle}{\emph{Proceedings of the 33rd Annual ACM Conference on
  Human Factors in Computing Systems}}. ACM, \bibinfo{publisher}{ACM},
  \bibinfo{address}{New York, NY, USA}, \bibinfo{pages}{135--144}.
\newblock


\bibitem[\protect\citeauthoryear{Janzen and Teather}{Janzen and
  Teather}{2014}]%
        {janzen201460}
\bibfield{author}{\bibinfo{person}{Benjamin~F Janzen} {and}
  \bibinfo{person}{Robert~J Teather}.} \bibinfo{year}{2014}\natexlab{}.
\newblock \showarticletitle{Is 60 FPS better than 30?: the impact of frame rate
  and latency on moving target selection}. In
  \bibinfo{booktitle}{\emph{Proceedings of the extended abstracts of the 32nd
  annual ACM conference on Human factors in computing systems}}. ACM,
  \bibinfo{publisher}{ACM}, \bibinfo{address}{New York, NY, USA},
  \bibinfo{pages}{1477--1482}.
\newblock


\bibitem[\protect\citeauthoryear{Jota, Ng, Dietz, and Wigdor}{Jota
  et~al\mbox{.}}{2013}]%
        {jota2013fast}
\bibfield{author}{\bibinfo{person}{Ricardo Jota}, \bibinfo{person}{Albert Ng},
  \bibinfo{person}{Paul Dietz}, {and} \bibinfo{person}{Daniel Wigdor}.}
  \bibinfo{year}{2013}\natexlab{}.
\newblock \showarticletitle{How fast is fast enough?: a study of the effects of
  latency in direct-touch pointing tasks}. In
  \bibinfo{booktitle}{\emph{Proceedings of the sigchi conference on human
  factors in computing systems}}. ACM, \bibinfo{publisher}{ACM},
  \bibinfo{address}{New York, NY, USA}, \bibinfo{pages}{2291--2300}.
\newblock


\bibitem[\protect\citeauthoryear{Kampman}{Kampman}{2017}]%
        {pcrefresh}
\bibfield{author}{\bibinfo{person}{J Kampman}.}
  \bibinfo{year}{2017}\natexlab{}.
\newblock \bibinfo{booktitle}{\emph{Poll: What's the resolution and refresh
  rate of your gaming monitor?}}
\newblock The Tech Report.
\newblock
\urldef\tempurl%
\url{https://techreport.com/news/31542/poll-whats-the-resolution-and-refresh-rate-of-your-gaming-monitor/}
\showURL{%
\tempurl}


\bibitem[\protect\citeauthoryear{Landis}{Landis}{1954}]%
        {landis1954determinants}
\bibfield{author}{\bibinfo{person}{Carney Landis}.}
  \bibinfo{year}{1954}\natexlab{}.
\newblock \showarticletitle{Determinants of the critical flicker-fusion
  threshold}.
\newblock \bibinfo{journal}{\emph{Physiological Reviews}} \bibinfo{volume}{34},
  \bibinfo{number}{2} (\bibinfo{year}{1954}), \bibinfo{pages}{259--286}.
\newblock


\bibitem[\protect\citeauthoryear{Lane, Banaji, Nosek, and Greenwald}{Lane
  et~al\mbox{.}}{2007}]%
        {lane2007understanding}
\bibfield{author}{\bibinfo{person}{Kristin~A Lane}, \bibinfo{person}{Mahzarin~R
  Banaji}, \bibinfo{person}{Brian~A Nosek}, {and} \bibinfo{person}{Anthony~G
  Greenwald}.} \bibinfo{year}{2007}\natexlab{}.
\newblock \showarticletitle{Understanding and using the implicit association
  test: IV}.
\newblock In \bibinfo{booktitle}{\emph{Implicit Measures of Attitudes}}.
  \bibinfo{publisher}{Guilford}, \bibinfo{address}{New York, NY USA},
  \bibinfo{pages}{59--102}.
\newblock


\bibitem[\protect\citeauthoryear{Lee and Chang}{Lee and Chang}{2018}]%
        {lee2018enhancing}
\bibfield{author}{\bibinfo{person}{Steven~WK Lee} {and}
  \bibinfo{person}{Rocky~KC Chang}.} \bibinfo{year}{2018}\natexlab{}.
\newblock \showarticletitle{Enhancing the experience of multiplayer shooter
  games via advanced lag compensation}. In
  \bibinfo{booktitle}{\emph{Proceedings of the 9th ACM Multimedia Systems
  Conference}}. ACM, \bibinfo{publisher}{ACM}, \bibinfo{address}{New York, NY,
  USA}, \bibinfo{pages}{284--293}.
\newblock


\bibitem[\protect\citeauthoryear{Liu, Agrawal, Sarkar, and Chen}{Liu
  et~al\mbox{.}}{2009}]%
        {liu2009dynamic}
\bibfield{author}{\bibinfo{person}{Changchun Liu}, \bibinfo{person}{Pramila
  Agrawal}, \bibinfo{person}{Nilanjan Sarkar}, {and} \bibinfo{person}{Shuo
  Chen}.} \bibinfo{year}{2009}\natexlab{}.
\newblock \showarticletitle{Dynamic difficulty adjustment in computer games
  through real-time anxiety-based affective feedback}.
\newblock \bibinfo{journal}{\emph{International Journal of Human-Computer
  Interaction}} \bibinfo{volume}{25}, \bibinfo{number}{6}
  (\bibinfo{year}{2009}), \bibinfo{pages}{506--529}.
\newblock


\bibitem[\protect\citeauthoryear{Long and Gutwin}{Long and Gutwin}{2018}]%
        {Long:2018:CME:3242671.3242678}
\bibfield{author}{\bibinfo{person}{Michael Long} {and} \bibinfo{person}{Carl
  Gutwin}.} \bibinfo{year}{2018}\natexlab{}.
\newblock \showarticletitle{Characterizing and Modeling the Effects of Local
  Latency on Game Performance and Experience}. In
  \bibinfo{booktitle}{\emph{Proceedings of the 2018 Annual Symposium on
  Computer-Human Interaction in Play}} \emph{(\bibinfo{series}{CHI PLAY '18})}.
  \bibinfo{publisher}{ACM}, \bibinfo{address}{New York, NY, USA},
  \bibinfo{pages}{285--297}.
\newblock
\showISBNx{978-1-4503-5624-4}
\urldef\tempurl%
\url{https://doi.org/10.1145/3242671.3242678}
\showDOI{\tempurl}


\bibitem[\protect\citeauthoryear{Long and Gutwin}{Long and Gutwin}{2019}]%
        {long2019effects}
\bibfield{author}{\bibinfo{person}{Michael Long} {and} \bibinfo{person}{Carl
  Gutwin}.} \bibinfo{year}{2019}\natexlab{}.
\newblock \showarticletitle{Effects of Local Latency on Game Pointing Devices
  and Game Pointing Tasks}. In \bibinfo{booktitle}{\emph{Proceedings of the
  2019 CHI Conference on Human Factors in Computing Systems}}. ACM,
  \bibinfo{publisher}{ACM}, \bibinfo{address}{New York, NY, USA},
  \bibinfo{pages}{208}.
\newblock


\bibitem[\protect\citeauthoryear{Meade}{Meade}{2009}]%
        {meade2009freeiat}
\bibfield{author}{\bibinfo{person}{Adam~W Meade}.}
  \bibinfo{year}{2009}\natexlab{}.
\newblock \showarticletitle{FreeIAT: An open-source program to administer the
  Implicit Association Test.}
\newblock \bibinfo{journal}{\emph{Applied psychological measurement}}
  \bibinfo{volume}{33}, \bibinfo{number}{8} (\bibinfo{year}{2009}),
  \bibinfo{pages}{643}.
\newblock
\urldef\tempurl%
\url{https://psycnet.apa.org/doi/10.1177/0146621608327803}
\showURL{%
\tempurl}


\bibitem[\protect\citeauthoryear{Mehrabian and Russell}{Mehrabian and
  Russell}{1974}]%
        {mehrabian1974approach}
\bibfield{author}{\bibinfo{person}{Albert Mehrabian} {and}
  \bibinfo{person}{James~A Russell}.} \bibinfo{year}{1974}\natexlab{}.
\newblock \bibinfo{booktitle}{\emph{An approach to environmental psychology.}}
\newblock \bibinfo{publisher}{MIT Press}, \bibinfo{address}{Cambridge, MA USA}.
\newblock


\bibitem[\protect\citeauthoryear{Nakamura and Csikszentmihalyi}{Nakamura and
  Csikszentmihalyi}{2014}]%
        {nakamura2014concept}
\bibfield{author}{\bibinfo{person}{Jeanne Nakamura} {and}
  \bibinfo{person}{Mihaly Csikszentmihalyi}.} \bibinfo{year}{2014}\natexlab{}.
\newblock \showarticletitle{The concept of flow}.
\newblock In \bibinfo{booktitle}{\emph{Flow and the foundations of positive
  psychology}}. \bibinfo{publisher}{Springer Verlag}, \bibinfo{address}{London,
  UK}, \bibinfo{pages}{239--263}.
\newblock


\bibitem[\protect\citeauthoryear{Newzoo}{Newzoo}{2019}]%
        {esportview2019}
\bibfield{author}{\bibinfo{person}{Newzoo}.} \bibinfo{year}{2019}\natexlab{}.
\newblock \bibinfo{booktitle}{\emph{New Twitch Rankings: Top Games by Esports
  and Total Viewing Hours}}.
\newblock Newzoo.
\newblock
\urldef\tempurl%
\url{https://newzoo.com/insights/articles/new-twitch-rankings-top-games-esports-total-viewing-hours/}
\showURL{%
\tempurl}


\bibitem[\protect\citeauthoryear{NVIDIA}{NVIDIA}{2019}]%
        {GSYNCUlt35:online}
\bibfield{author}{\bibinfo{person}{NVIDIA}.} \bibinfo{year}{2019}\natexlab{}.
\newblock \bibinfo{booktitle}{\emph{G-SYNC Ultimate Gaming Monitors}}.
\newblock NVIDIA.
\newblock
\urldef\tempurl%
\url{https://www.nvidia.com/en-us/geforce/products/g-sync-monitors/}
\showURL{%
\tempurl}


\bibitem[\protect\citeauthoryear{Pannekeet}{Pannekeet}{2019}]%
        {esport2019}
\bibfield{author}{\bibinfo{person}{Jurre Pannekeet}.}
  \bibinfo{year}{2019}\natexlab{}.
\newblock \bibinfo{booktitle}{\emph{Global Esports Economy Will Top \$ 1
  Billion for the First Time in 2019}}.
\newblock Newzoo.
\newblock
\urldef\tempurl%
\url{https://newzoo.com/insights/articles/newzoo-global-esports-economy-will-top-1-billion-for-the-first-time-in-2019/}
\showURL{%
\tempurl}


\bibitem[\protect\citeauthoryear{Poth, Foerster, Behler, Schwanecke, Schneider,
  and Botsch}{Poth et~al\mbox{.}}{2018}]%
        {poth2018ultrahigh}
\bibfield{author}{\bibinfo{person}{Christian~H Poth},
  \bibinfo{person}{Rebecca~M Foerster}, \bibinfo{person}{Christian Behler},
  \bibinfo{person}{Ulrich Schwanecke}, \bibinfo{person}{Werner~X Schneider},
  {and} \bibinfo{person}{Mario Botsch}.} \bibinfo{year}{2018}\natexlab{}.
\newblock \showarticletitle{Ultrahigh temporal resolution of visual
  presentation using gaming monitors and G-Sync}.
\newblock \bibinfo{journal}{\emph{Behavior research methods}}
  \bibinfo{volume}{50}, \bibinfo{number}{1} (\bibinfo{year}{2018}),
  \bibinfo{pages}{26--38}.
\newblock


\bibitem[\protect\citeauthoryear{Roohi, Mekler, Tavast, Blomqvist, and
  H{\"a}m{\"a}l{\"a}inen}{Roohi et~al\mbox{.}}{2019}]%
        {roohi2019recognizing}
\bibfield{author}{\bibinfo{person}{Shaghayegh Roohi}, \bibinfo{person}{Elisa~D
  Mekler}, \bibinfo{person}{Mikke Tavast}, \bibinfo{person}{Tatu Blomqvist},
  {and} \bibinfo{person}{Perttu H{\"a}m{\"a}l{\"a}inen}.}
  \bibinfo{year}{2019}\natexlab{}.
\newblock \showarticletitle{Recognizing Emotional Expression in Game Streams}.
  In \bibinfo{booktitle}{\emph{Proceedings of the Annual Symposium on
  Computer-Human Interaction in Play}}. \bibinfo{publisher}{ACM},
  \bibinfo{address}{New York, NY, USA}, \bibinfo{pages}{301--311}.
\newblock


\bibitem[\protect\citeauthoryear{Sarkar}{Sarkar}{2014}]%
        {sarkar2014frame}
\bibfield{author}{\bibinfo{person}{Samit Sarkar}.}
  \bibinfo{year}{2014}\natexlab{}.
\newblock \bibinfo{booktitle}{\emph{Why frame rate and resolution matter: A
  graphics primer}}.
\newblock Polygon.
\newblock
\urldef\tempurl%
\url{https://www.polygon.com/2014/6/5/5761780/frame-rate-resolution-graphics-primer-ps4-xbox-one}
\showURL{%
\tempurl}


\bibitem[\protect\citeauthoryear{Spjut, Boudaoud, Binaee, Kim, Majercik,
  McGuire, Luebke, and Kim}{Spjut et~al\mbox{.}}{2019}]%
        {spjut2019latency}
\bibfield{author}{\bibinfo{person}{Josef Spjut}, \bibinfo{person}{Ben
  Boudaoud}, \bibinfo{person}{Kamran Binaee}, \bibinfo{person}{Jonghyun Kim},
  \bibinfo{person}{Alexander Majercik}, \bibinfo{person}{Morgan McGuire},
  \bibinfo{person}{David Luebke}, {and} \bibinfo{person}{Joohwan Kim}.}
  \bibinfo{year}{2019}\natexlab{}.
\newblock \showarticletitle{Latency of 30 ms Benefits First Person Targeting
  Tasks More Than Refresh Rate Above 60 Hz}. In
  \bibinfo{booktitle}{\emph{SIGGRAPH Asia 2019 Technical Briefs}}.
  \bibinfo{publisher}{ACM}, \bibinfo{address}{New York, NY, USA},
  \bibinfo{pages}{110--113}.
\newblock


\bibitem[\protect\citeauthoryear{Stewart}{Stewart}{2019}]%
        {gamingscan}
\bibfield{author}{\bibinfo{person}{Samuel Stewart}.}
  \bibinfo{year}{2019}\natexlab{}.
\newblock \bibinfo{booktitle}{\emph{What Is The Best FPS For Gaming?}}
\newblock Gaming Scan.
\newblock
\urldef\tempurl%
\url{https://www.gamingscan.com/best-fps-gaming/}
\showURL{%
\tempurl}


\bibitem[\protect\citeauthoryear{Watson and Ahumada}{Watson and
  Ahumada}{2011}]%
        {watson201164}
\bibfield{author}{\bibinfo{person}{Andrew~B Watson} {and}
  \bibinfo{person}{Albert~J Ahumada}.} \bibinfo{year}{2011}\natexlab{}.
\newblock \showarticletitle{64.3: flicker visibility: a perceptual metric for
  display flicker}. In \bibinfo{booktitle}{\emph{SID symposium digest of
  Technical Papers}}, Vol.~\bibinfo{volume}{42}. \bibinfo{publisher}{Wiley
  Online Library}, \bibinfo{address}{Hoboken, NJ USA},
  \bibinfo{pages}{957--959}.
\newblock


\bibitem[\protect\citeauthoryear{Watson, Shrivastava, and Gavane}{Watson
  et~al\mbox{.}}{2019}]%
        {watson2019effects}
\bibfield{author}{\bibinfo{person}{Benjamin Watson}, \bibinfo{person}{Rachit
  Shrivastava}, {and} \bibinfo{person}{Ajinkya Gavane}.}
  \bibinfo{year}{2019}\natexlab{}.
\newblock \showarticletitle{The Effects of Adaptive Synchronization on
  Performance and Experience in Gameplay}.
\newblock \bibinfo{journal}{\emph{Proceedings of the ACM on Computer Graphics
  and Interactive Techniques}} \bibinfo{volume}{2}, \bibinfo{number}{1}
  (\bibinfo{year}{2019}), \bibinfo{pages}{5}.
\newblock


\bibitem[\protect\citeauthoryear{Wattimena, Kooij, Van~Vugt, and
  Ahmed}{Wattimena et~al\mbox{.}}{2006}]%
        {wattimena2006predicting}
\bibfield{author}{\bibinfo{person}{AF Wattimena}, \bibinfo{person}{Robert~E
  Kooij}, \bibinfo{person}{JM Van~Vugt}, {and} \bibinfo{person}{OK Ahmed}.}
  \bibinfo{year}{2006}\natexlab{}.
\newblock \showarticletitle{Predicting the perceived quality of a first person
  shooter: the Quake IV G-model}. In \bibinfo{booktitle}{\emph{Proceedings of
  5th ACM SIGCOMM workshop on Network and system support for games}}. ACM,
  \bibinfo{publisher}{ACM}, \bibinfo{address}{New York, NY, USA},
  \bibinfo{pages}{42}.
\newblock


\bibitem[\protect\citeauthoryear{Wiebe, Lamb, Hardy, and Sharek}{Wiebe
  et~al\mbox{.}}{2014}]%
        {wiebe2014measuring}
\bibfield{author}{\bibinfo{person}{Eric~N Wiebe}, \bibinfo{person}{Allison
  Lamb}, \bibinfo{person}{Megan Hardy}, {and} \bibinfo{person}{David Sharek}.}
  \bibinfo{year}{2014}\natexlab{}.
\newblock \showarticletitle{Measuring engagement in video game-based
  environments: Investigation of the User Engagement Scale}.
\newblock \bibinfo{journal}{\emph{Computers in Human Behavior}}
  \bibinfo{volume}{32} (\bibinfo{year}{2014}), \bibinfo{pages}{123--132}.
\newblock


\end{thebibliography}

\end{document}